\documentclass{sclass}

\numberwithin{equation}{section}

\begin{document}

\renewcommand{\PaperNumber}{***}

\FirstPageHeading

\ShortArticleName{An Algebraic Model for the Multiple Meixner Polynomials of the First Kind}

\ArticleName{An Algebraic Model for the Multiple Meixner \\ Polynomials of the First Kind}

\Author{Hiroshi Miki~$^\dag$, Satoshi Tsujimoto~$^\dag$, Luc Vinet~$^\ddag$ and Alexei Zhedanov~$^\S$}

\AuthorNameForHeading{H Miki, S. Tsujimoto, L. Vinet and A. Zhedanov}

\Address{$^\dag$~Department of Applied Mathematics and Physics, Graduate School of Informatics, Kyoto University, Sakyo-Ku, Kyoto 606 8501, Japan} 
\EmailD{miki@amp.i.kyoto-u.ac.jp,tujimoto@i.kyoto-u.ac.jp} 

\Address{$^\ddag$~Centre de recherches math\'{e}matiques, Universit\'{e} de Montr\'{e}al, P. O. Box 6128, Centre-ville Station, Montr\'{e}al (Qu\'{e}bec), H3C 3J7, Canada}
\EmailD{luc.vinet@umontreal.ca} 

\Address{$^\S$~Donetsk Institute for Physics and Technology, Donetsk 83 114, Ukraine}
\EmailD{zhedanov@fti.dn.ua}


\Abstract{An interpretation of the multiple Meixner polynomials of the first kind is provided through an infinite Lie algebra realized in terms of the creation and annihilation operators of a set of independent oscillators. The model is used to derive properties of these orthogonal polynomials.}


\Classification{20C35; 42C05; 33C80; 81R05; 81R30} 

\section{Introduction}
The theory of multiple orthogonal polynomials, rooted in Hermite-Pad\'{e} approximation schemes \cite{Angelesco,Nikishin}, is currently undergoing development \cite{Assche,Kuijlaars,Haneczok,Ismail}.
These are polynomials of a single variable, labeled by a multi-index and orthogonal for a number of measures. Three of us have recently shown \cite{Miki} that the multiple Charlier polynomials arise in the construction of the common eigenstates of a set of non-Hermitian oscillator Hamiltonians. In keeping with the significant connection between algebras and special functions, this provided a model from which many properties of the multiple Charlier polynomials could be derived and interpreted while giving a physical context for them. The purpose of this paper is to similarly describe another family of multiple orthogonal polynomials, namely that of the multiple Meixner polynomials of the first kind. As shall be seen, these  are related to the Lie algebra of diffeomorphisms in $\mathbb{C}^r$ whose generators are realized in terms of the creation and annihilation oscillators of independent harmonic oscillators. This will allow to obtain their step relations and their generating function.

\section{Multiple Meixner polynomials of the first kind}
The monic $r$-multiple Meixner polynomials of the first kind $M_{\vec{n}}^{\beta, \vec{c}}(x)$ are labeled by a multi-index $\vec{n}=(n_1,\cdots, n_r)\in \mathbb{N}^r$ with length $|\vec{n}|=n_1+\cdots +n_r$. They are orthogonal with respect to the $r$ discrete measures
\begin{equation}
\mu _i=\sum_{x=0}^{\infty } (\beta)_x \frac{c_i^x}{x!}\delta_x ,\quad i=1,\cdots ,r,
\end{equation} 
where $\beta >0 ,0< c_i<1$ for all $i$ and $c_i\ne c_j$ whenever $i\ne j$; that is, they satisfy
\begin{equation}
\sum_{x=0}^{\infty }M_{\vec{n}}^{\beta ,\vec{c}}(x)x^j(\beta )_x\frac{c_i^x}{x!}=0,
\end{equation}
for $j=0,\cdots ,n_i-1,i=1,\cdots ,r$. As usual, the Pochhammer symbol stands for $(\beta)_x=\beta (\beta +1)\cdots (\beta +x-1)$. The parameters $c_1,\cdots ,c_r$ are collectively represented by $\vec{c}$. In the following, we shall often omit the suffices $\beta $ and $\vec{c}$ whenever the context is unambiguous.

Let $\vec{e}_i=(0,\cdots ,0,1,0,\cdots ,0)$ denote the $i$-th standard unit vector in $r$ dimensions with $1$ in the $i$-th entry. The multiple Meixner polynomials of the first kind obey \cite{Haneczok} $r$ recurrence relations of nearest-neighbor form that read as follows:
\begin{align}\label{nearest-neighbor}
\begin{split}
xM_{\vec{n}}(x)
&=M_{\vec{n}+\vec{e}_i}(x)+\left[ \frac{c_i}{1-c_i}(\beta +|\vec{n}|)+\sum_{k=1}^r \frac{n_k}{1-c_k}\right] M_{\vec{n}}(x)\\
&+\sum_{k=1}^r \frac{c_k}{(1-c_k)^2}n_k(\beta +|\vec{n}|-1)M_{\vec{n}-\vec{e}_k}(x),\quad i=1,\cdots ,r.
\end{split}
\end{align}
It should be observed, that if one subtracts the above relations pair-wise, one finds that the polynomials $M_{\vec{n}}(x)$ verify as a consequence of  \eqref{nearest-neighbor}:
\begin{equation}\label{commuting}
M_{\vec{n}+\vec{e}_i}(x)-M_{\vec{n}+\vec{e}_j}=(\beta+|\vec{n}|)\frac{c_j-c_i}{(1-c_i)(1-c_j)}M_{\vec{n}}(x)
\end{equation}
for all $i,j\in \{ 1,\cdots ,r\}$.

\section{Harmonic oscillator operators and states}
The construction of the algebraic model for the polynomials $M_{\vec{n}}^{\beta, \vec{c}}$ will involve the operators and eigenstates of $r$ independent quantum harmonic oscillators . 
We here briefly review basic facts in this connection to establish notation.

The annihilation and creation operators $a_i$, $a_i^+$ (resp.), $i=1,\cdots ,r$, satisfy the commutation relations
\begin{equation}
[a_i,a_j]=[a_i^+,a_j^+]=0,\quad [a_i,a_j^+]=\delta_{i,j},\quad i,j=1,\cdots, r
\end{equation}
and hence generate the Lie algebra $W(r)$ made out of $r$ copies $W_i$ of the Heisenberg-Weyl algebra:
\begin{equation}
W(r)=\bigoplus_{i=1}^r W_i,
\end{equation}
where $W_i$ is the linear span of $\{ a_i,a_i^+,1\}$. Denote by $\left| n_1,\cdots ,n_r\right>=\left|n_1\right> \cdots \left| n_r\right>$ the normalized simultaneous eigenvectors of the $r$ number operators $N_i=a_i^+a_i$:
\begin{align}
&a_i^+a_i\left| n_1,\cdots,n_i,\cdots  ,n_r\right>=n_i\left| n_1,\cdots,n_i,\cdots  ,n_r\right>, \quad n_i\in \mathbb{N},\quad i=1,\cdots ,r, \\
&\left< m_1,\cdots ,m_r|n_1,\cdots ,n_r\right>=\delta_{m_1,n_1}\cdots \delta_{m_r, n_r}.
\end{align} 
The algebra $W(r)$ is represented in the standard way in this number state basis:
\begin{align} \label{aa_action}
\begin{split}
a_i\left| n_1,\cdots ,n_i,\cdots ,n_r\right>&=\sqrt{n_i} \left| n_1,\cdots ,n_i-1,\cdots ,n_r\right>,\\
a_i^+\left| n_1,\cdots ,n_i,\cdots ,n_r\right>&=\sqrt{n_i+1}\left| n_1,\cdots ,n_i+1,\cdots ,n_r\right>.
\end{split}
\end{align}
The Bargmann realization in terms of coordinates $z_i,i=1,\cdots ,r$ in $\mathbb{C}^r$ has 
\begin{equation}\label{Bargmann}
a_i=\frac{\partial}{\partial z_i},\quad a_i^+=z_i ,\quad i=1,\cdots, r, ~~\textrm{and }~~\left< z_1,\cdots ,z_r | n_1,\cdots ,n_r \right> = \frac{z_1^{n_1}\cdots z_r^{n_r}}{\sqrt{n_1!\cdots n_r!}}.
\end{equation}

\section{The Model}\label{model}
Consider the set of $r$ non-Hermitian operators $H_i,i=1,.\cdots ,r$, defined as follows in $\mathcal{U}(W(r))$, the universal enveloping algebra of $W(r)$: 
\begin{align}
\begin{split}
H_i^{\beta,\vec{c}} = a_i+\sum_{k=1}^r \frac{a_k^+a_k}{1-c_k}+\left( \frac{c_i}{1-c_i}+\sum_{j=1}^r \frac{c_j}{(1-c_j)^2}a_j^+\right) \left( \sum_{k=1}^r a_k^+a_k +\beta \right) ,\quad i=1,\cdots ,r.
\end{split}
\end{align}
Let us observe, especially using the holomorphic representation of Bargmann, that the operators $\sum_{k=1}^r a_k^+a_k,a_i,a_i^+,a_i^+\sum_{k=1}^ra_k^+a_k,i=1,\cdots ,r,$ making up the $H_i$, generate a Lie algebra isomorphic to that of the diffeomorphisms in $\mathbb{C}^r$ spanned by (generalized) vector fields of the form $\mathcal{Z}=\sum_{i=1}^r f_i(\vec{z})\frac{\partial}{\partial z_i}+g(\vec{z})$ with $\vec{z}=(z_1,\cdots ,z_r)$.
In the coordinate realization where $a_i=\frac{1}{\sqrt{2}}(x_i+\frac{\partial}{\partial x_i})$ and $a_i^+=\frac{1}{\sqrt{2}}(x_i-\frac{\partial}{\partial x_i}),~i=1,\cdots ,r$, the operators $H_i$ are third order differential operators. 
However, they can be considered as Hamiltonians, as we shall see below. Now the key observation is that they can be simultaneously diagonalized by the multiple Meixner polynomials.

Indeed, consider the states $\left| \left. x, \beta ,\vec{c}~\right> \right>$ defined through the following combination of the states $\left| n_1,\cdots ,n_r\right>$:
\begin{equation}
\left| \left. x, \beta ,\vec{c}~\right> \right> = N_{x,\beta ,\vec{c}}^r \sum_{\vec{n}} \frac{M_{\vec{n}}^{\beta ,\vec{c}}(x)}{\sqrt{n_1!\cdots n_r!}} \left| n_1,\cdots ,n_r\right>,
\end{equation}
where $N_{x,\beta ,\vec{c}}^r$ are constants. Acting with $H_i$ on $\left| \left. x,\beta \right>\right>$, we find (with some indices suppressed):
\begin{align}
\begin{split}
H_i^{\beta }\left| \left. x,\beta \right>\right> 
&= N_x^r \sum_{\vec{n}=\vec{0}}^{\infty } \frac{M_{\vec{n}}(x)}{\sqrt{n_1!\cdots n_r!}} \cdot \biggl\{ \sqrt{n_i} \left| n_1,\cdots n_i-1,\cdots ,n_r\right>\\
&\quad +\left[ \sum_{k=1}^r \frac{n_k}{1-c_k}+\frac{c_i}{1-c_i}\left( \beta + \sum_{k=1}^r n_k\right) \right]\left| n_1,\cdots ,n_r\right> \\
&\quad +\left( \beta +\sum_{k=1}^rn_k\right) \sum_{j=1}^r \frac{c_j}{(1-c_j)^2} \sqrt{n_j+1}\left| n_1,\cdots ,n_i+1,\cdots ,n_r\right> \biggr\} \\
&=N_x^r \sum_{\vec{n}=\vec{0}}^{\infty } \frac{1}{\sqrt{n_1!\cdots n_r!}} \cdot \biggl\{ M_{\vec{n}+\vec{e}_i}(x)+\left[ \frac{c_i}{1-c_i}(\beta +|\vec{n}|)+ \sum_{k=1}^r \frac{n_k}{1-c_k}\right] M_{\vec{n}}(x)\\
&\quad +\sum_{j=1}^r \frac{c_j}{(1-c_j)^2}n_j (\beta +|\vec{n}|-1)M_{\vec{n}-\vec{e}_j}(x)\biggr\} \left| n_1,\cdots ,n_r\right>.
\end{split}
\end{align}
Hence, appealing to the recurrence relations \eqref{nearest-neighbor}, we have
\begin{equation}
H_i^{\beta ,\vec{c}} \left| \left. x, \beta ,\vec{c}~\right> \right> = x \left| \left. x, \beta ,\vec{c}~\right> \right>
\end{equation}
for all $i=1,\cdots ,r$. We thus find that in spite of being non-Hermitian, the operators $H_i^{\beta, \vec{c}}$ all have the same real spectrum. (An explanation is given below.) 


Somewhat surprisingly, the operators $H_i^{\beta, \vec{c}}$ do not commute. In effect, one finds that 
\begin{equation}
\left[ H_i^{\beta, \vec{c}}, H_j^{\beta, \vec{c}}\right] = a_i-a_j+ \frac{c_i-c_j}{(1-c_i)(1-c_j)}\left( \beta +\sum_{k=1}^r a_k^+a_k\right).
\end{equation}
How can this be reconciled with the fact that they can be simultaneously diagonalized? In this respect, one observes that
\begin{align}
\begin{split}
&\left[ H_i^{\beta, \vec{c}}, H_j^{\beta, \vec{c}}\right]  \left| \left. x, \beta ,\vec{c}~\right> \right> \\
&= N_x^r \sum_{\vec{n}=\vec{0}}^{\infty } \frac{M_{\vec{n}}^{\beta ,\vec{c}}(x)}{\sqrt{n_1!\cdots ,n_r!}}\cdot \biggl\{ \sqrt{n_i} \left| n_1,\cdots ,n_i-1,\cdots ,n_r\right>-\sqrt{n_j} \left| n_1,\cdots ,n_j-1,\cdots ,n_r\right> \\
&\quad + \frac{c_i-c_j}{(1-c_i)(1-c_j)}\left( \beta + |\vec{n}|\right)\sqrt{n_i} \left| n_1,\cdots ,n_r\right>  \biggr\}\\
&=N_x^r \sum_{\vec{n=0}}^{\infty } \left[ M_{\vec{n}+\vec{e}_i}(x)-M_{\vec{n}+\vec{e}_j}+\frac{c_i-c_j}{(1-c_i)(1-c_j)}\left( \beta +|\vec{n}|\right)M_{\vec{n}}(x)\right] \left| n_1,\cdots ,n_r\right> 
\end{split}
\end{align}
to find that 
\begin{equation}
\left[ H_i^{\beta, \vec{c}}, H_j^{\beta, \vec{c}}\right]  \left| \left. x, \beta ,\vec{c}~\right> \right> = 0
\end{equation}
from \eqref{commuting}. This shows that the operators $H_i^{\beta ,\vec{c}}$ commute when acting on the states $\left| \left. x, \beta ,\vec{c}~\right> \right>$. These operators therefore define a ``weakly'' integrable model.

\section{Conjugation operators and isospectrality}
First observe, using \eqref{commuting} in \eqref{nearest-neighbor}, that $M_{\vec{n}}(x)$ satisfies ``non nearest-neighbor'' recurrence relations of the form
\begin{align}\label{non-nearest-neighbor}
\begin{split}
xM_{\vec{n}}(x)
&=M_{\vec{n}+\vec{e}_i}(x)+\left( \frac{c_i}{1-c_i}\beta + \frac{|\vec{n}|}{1-c_i}\right) M_{\vec{n}}(x)+\sum_{k=i} \frac{c_k}{1-c_k}n_k M_{\vec{n}+\vec{e}_i-\vec{e}_k}(x) \\
&+ \sum_{k=1}^r\frac{c_k}{(1-c_i)(1-c_k)}n_k(\beta +|\vec{n}|-1)M_{\vec{n}-\vec{e}_k},\quad i=1,\cdots ,r.
\end{split}
\end{align}  
Consider another set of $r$ Hamiltonians $\bar{H}_i^{\beta ,\vec{c}}$ defined as follows:
\begin{align}
\begin{split}
\bar{H}_{i}^{\beta ,\vec{c}}
&= a_i+\frac{c_i}{1-c_i}\beta +\frac{\sum_{k=1}^{r} a_k^+a_k}{1-c_i}+\sum_{k=1}^r \frac{c_k}{1-c_k}a_k^+a_i \\
&+ \sum_{k=1}^r\frac{c_k}{(1-c_i)(1-c_k)}a_k^+\left( \beta +\sum_{k=1}^ra_k^+a_k\right) ,\quad i=1,\cdots ,r.
\end{split}
\end{align}  
In a manner similar to the procedure followed in Section \ref{model}, one can easily show that the operators $\bar{H}_i^{\beta, \vec{c}},i=1,\cdots ,r$ are also diagonalized with the help of the multiple Meixner polynomials.
In fact one finds that they share with $H_i^{\beta ,\vec{c}}$ the same eigenstates:
\begin{equation}
\bar{H}_{i}^{\beta ,\vec{c}}\left| \left. x, \beta ,\vec{c}~\right> \right> = x\left| \left. x, \beta ,\vec{c}~\right> \right>.
\end{equation}  
Here it should be noted that $\bar{H}_i^{\beta ,\vec{c}} \ne H_i^{\beta ,\vec{c}}$ except for $r=1$ , although they have common spectra by construction. 

Quite remarkably, each of the operators $\bar{H}_i^{\beta ,\vec{c}}$ can be obtained from the standard harmonic oscillator Hamiltonians in $r$ dimensions
\begin{equation}
H_0=\sum_{j=1}^r a_j^+a_j
\end{equation}
by a conjugation. Let $S_i,~i=1,\cdots ,r$ be the operators defined by
\begin{equation}
S_i=\exp\left[ (1-c_i)a_i\right] \prod_{j=1}^r \exp \left[ -\frac{c_k}{(1-c_i)(1-c_k)}a_k^+(\beta +H_0)\right].
\end{equation}
Using the Baker-Campbell-Hausdorff formula
\begin{equation}\label{bch}
\exp[X]~Y\exp[-X]=Y+[X,Y]+\frac{1}{2}[X,[X,Y]]+\cdots 
\end{equation}
and the fact that
\begin{subequations}
\begin{align}
&[a_k^+(\beta +H_0),H_0]=a_k^+(\beta +H_0),\\
&[a_i^+(\beta +H_0),a_j^+(\beta +H_0)]=0,
\end{align}
\end{subequations}
it is seen that
\begin{equation}
\bar{H}_i^{\beta ,\vec{c}}=S_iH_0S_i^{-1}
\end{equation}
with the proper choice for $\beta $ and $\vec{c}$ in $S_i$ implicit.
This explains why each of the operators $\bar{H}_i^{\beta ,\vec{c}}$ and hence each of the $H_i^{\beta ,\vec{c}}$ has  a real spectrum identical to that of $H_0$.

\section{The special case $r=1$ and $SU(1,1)$}

When $r=1$, the multiple Meixner polynomials of the first kind reduce to the standard monic Meixner polynomials. It is well known that these are related to the discrete series representation of the Lie algebra $SU(1,1)$ which is bounded from below \cite{Granovskii,Floreanini,Vinet}. It is of interest to make the connection with our model.

Let $J_0$ and $J_{\pm}$ denote the generators of $SU(1,1)$ satisfying the commutation relations:
\begin{equation}\label{deforming}
[J_0,J{\pm}] = \pm J_{\pm},\quad [J_+,J_-]=-2J_0.
\end{equation}
The Casimir operator $C$ is given by
\begin{equation}
C=J_0^2-\frac{1}{2}\left( J_+J_-+J_-J_+\right).
\end{equation}
We shall take it to be of the following form
\begin{equation}
C=d(d-1)
\end{equation}
in irreducible representations. The discrete series model of $SU(1,1)$ which is bounded from below is spanned by basis vectors $\{ \left| d, n\right>,n=0,1,\cdots \}$ on which the generators act according to \cite{Wybourne}
\begin{subequations}
\begin{align}
J_0 \left| d,n\right> &= (d+n) J_0 \left| d,n\right>,\\
J_{\pm} \left| d,n\right> &=\sqrt{(d+n\pm 1)(d+n)-d(d-1)} \left| d,n\pm 1\right> . 
\end{align}
\end{subequations}
Consider now the following ``Hamiltonians'' $\widetilde{H}$ in the Lie algebra of $SU(1,1)$:
\begin{equation}\label{su11hamiltonian}
\widetilde{H}=J_-+\left( \frac{1+c}{1-c}\right) J_0+\frac{c}{(1-c)^2}J_+, \quad 0<c<1.
\end{equation}
Using the Baker-Campbell-Hausdorff formula \eqref{bch}, it is readily seen that
\begin{equation}
\widetilde{H}=SJ_0S^{-1}
\end{equation}
with 
\begin{equation}\label{su11conjugate}
S=\exp [(1-c)J_-] \exp \left[ -\frac{c}{(1-c)^2}J_+\right]. 
\end{equation}
Hence $\widetilde{H}$ has the same spectrum as $J_0$. In order to diagonalize $\widetilde{H}$, one introduces the vectors
\begin{equation}
\left| d,x\right> = \sum_{n=0}^{\infty } \frac{p_n(x)}{\sqrt{n!(2d )_n}}\left| d,n\right>.
\end{equation}
The eigenvalue equation
\begin{equation}
\widetilde{H} \left| d,x\right> = (d+x) \left| d,x\right>
\end{equation} 
is then readily seen to require that the $p_n(x)$ satisfy the 3-term recurrence relation which is obtained from \eqref{nearest-neighbor} when $r=1$, as long as  
\begin{equation}
\beta = 2d.
\end{equation}
This establishes the connection between $SU(1,1)$ and the standard Meixner polynomials. 

Obtaining harmonic oscillator models for these polynomials therefore amounts to providing embeddings of $SU(1,1)$ in the universal enveloping algebra of $W(1)$. One such well known realization corresponds to the metaplectic representation of $SU(1,1)$. One takes
\begin{equation}\label{embed}
J_-=\frac{a^2}{2},\quad J_+=\frac{(a^+)^2}{2},\quad J_0=\frac{1}{4}(aa^++a^+a).
\end{equation}
It is straightforward to check that the defining relations \eqref{deforming} are satisfied, given the canonical commutation rule $[a,a^+]=1$. In this realization, the Casimir operator $C$ takes the value
\begin{equation}
C=d(d-1)=-\frac{3}{16},
\end{equation}
implying that only two series, respectively corresponding to $d=\frac{1}{4}$ and $d=\frac{3}{4}$ are thus constructed. In other words, this only provides an algebraic model for the standard Meixner polynomials $M_{n}^{\beta ,c}$ with $\beta = \frac{1}{2}$ or $\beta =\frac{3}{2}$. An extension to more general value of $\beta $ can be obtained \cite{Tsujimoto} by letting $a$ and $a^+$ in \eqref{embed} stand for annihilation and creation operators of parabosonic oscillators, that is by considering  a deformation of the Heisenberg algebra defined by the commutation relation 
\begin{equation}
[a,a^+]=1+2\mu R
\end{equation}
where $R$ is  an involution operator satisfying $R^2=1$.
In doing so, we construct the $SU(1,1)$ discrete series with $d( =\frac{\beta }{2})$ equal to $\frac{\mu}{2}+\frac{1}{4}$ and to $\frac{\mu}{2}+\frac{3}{4}$ 
and models for Meixner polynomials with the corresponding $\beta $.
In any case, we have found that the scheme corresponding to \eqref{embed} does not extend to $r$ dimensions or to $r$ oscillators, so as to provide an interpretation of the multiple Meixner polynomials.

Now, if we specialize the model presented in section \ref{model}, to the case $r=1$ (and suppress the sole index $1$), we first find that the operators $H$ and $\bar{H}$ coincide and read
\begin{equation}\label{embed2}
H=\bar{H}=a+\frac{1+c}{1-c}\left( a^+a+\frac{\beta }{2}\right) + \frac{c}{(1-c)^2}[a^+(\beta +a^+a)]-\frac{\beta }{2}.
\end{equation}
Moreover the operator that conjugates $H_0=a^+a$ into $\bar{H}=H$ is 
\begin{equation}
S=\exp[(1-c)a] \exp \left[ -\frac{c}{(1-c)^2}a^+(\beta +a^+a)\right] .
\end{equation}
Hence if we take
\begin{equation}
J_-=a,\quad J_+=a^+(\beta +a^+a),\quad J_0=a^+a+\frac{\beta }{2},
\end{equation}
we see that $H+\frac{\beta }{2}$ is in the form \eqref{su11hamiltonian} and that $S$ is the same group element as in \eqref{su11conjugate}.
It is then straightforward to check that the operators defined in \eqref{embed2} indeed verify the $SU(1,1)$ algebra commutation relations. Furthermore, the Casimir operator $C$ is readily calculated to be
\begin{equation}
C=\frac{\beta }{2}\left( \frac{\beta }{2}-1\right),
\end{equation} 
confirming that $d=\frac{\beta }{2}$ and that this embedding of $SU(1,1)$ in $\mathcal{U}(W(1))$ provides an algebraic model for the Meixner polynomials with arbitrary values of $\beta >0$. To our knowledge, this interpretation of the standard Meixner polynomials through the oscillator realization \eqref{embed2} of $SU(1,1)$ is new. The derivations of properties of the multiple Meixner polynomials stemming from our model would hence also be original in the case $r=1$. (See the next three sections.) Of course, for $r>1$, the multi-oscillator generalization of the operators \eqref{embed2} leads to generators that realize an infinite Lie algebra isomorphic to $\mathrm{Diff}(\mathbb{C}^r)$ as we have already pointed out.

\section{Ladder operators and step relations}
In the remaining sections, we shall illustrate how the algebraic model, presented in section \ref{model}, provides a useful framework to derive properties of the multiple Meixner polynomials. We will first focus on the step relations and therefore start with an examination of ladder operators.

Introduce 
\begin{equation}\label{raising1}
X_i=a_i+\frac{c_i}{1-c_i}(\beta +H_0),\quad i=1,\cdots ,r
\end{equation}
and
\begin{equation}\label{ladder1}
Y=1+\sum_{k=1}^r \frac{a_k^+}{1-c_k}.
\end{equation}
 It is straightforward to check that these operators satisfy the following relations:
 \begin{equation}\label{raising1b}
 H_i^{\beta +1}X_j-X_jH_i^{\beta }=-X_j
 \end{equation}
 and
 \begin{equation}\label{ladder1b}
 H_i^{\beta -1}Y-YH_i^{\beta }=Y,
 \end{equation}
 for $i,j=1,\cdots ,r$. From \eqref{raising1b} and \eqref{ladder1b}, it follows that
 \begin{align}
 X_i \left| \left. x,\beta \right> \right> &= \xi _{i,x,\beta } \left| \left. x-1,\beta +1 \right> \right>,\label{raising1a}\\
 Y \left| \left. x,\beta \right> \right> &= \eta _{x,\beta } \left| \left. x+1,\beta -1 \right> \right>\label{ladder1a},
 \end{align}
 where $\xi _{i,x,\beta }$ and $\eta _{x,\beta }$ are proportionality factors to be determined. Therefore $X_i$ (resp. $Y$) maps the eigenstate of $H_i^{\beta }$ with eigenvalue $x$ to the eigenstate of $H_i^{\beta +1}$ (resp. $H_i^{\beta-1}$) with eigenvalue $x+1$ (resp. $x-1$). It should be stressed that these operators (much like supersymmetry generators in quantum mechanics) relate the eigenstates of different operators.  Using the definition of $\left| \left. x,\beta \right> \right> $
 and a procedure by now familiar, it is found that \eqref{raising1a} translates into 
 \begin{equation}
 N_{x,\beta}^r\left[ M_{\vec{n}+\vec{e}_i}^{\beta }(x)+\frac{c_i}{1-c_i}(\beta +|\vec{n}|)M_{\vec{n}}^{\beta }(x)\right] = \xi_{i,x,\beta }N_{x-1,\beta +1}^{r} M_{\vec{n}}^{\beta +1}(x-1).
 \end{equation}
 Given the monicity of the polynomials, i.e. that the coefficient of the leading term is one, we must have
 \begin{equation}
 \xi_{i,x,\beta }= \frac{N_{x,\beta }^r}{N_{x-1,\beta +1}^r}(x+a),
 \end{equation}
 where $a$ is a constant. This gives
 \begin{equation}
M_{\vec{n}+\vec{e}_i}^{\beta }(x)+\frac{c_i}{1-c_i}(\beta +|\vec{n}|)M_{\vec{n}}^{\beta }(x)= (x+a)M_{\vec{n}}^{\beta +1}(x-1).
 \end{equation}
 In order to determine the constant $a$, we set $\vec{n}=\vec{0}$. We immediately find that $a=0$ since we know from the recurrence relations \eqref{nearest-neighbor} that
 \begin{equation}
 M_{\vec{e}_i}^{\beta ,\vec{c}}=x-\frac{c_i}{1-c_i}\beta .
 \end{equation}
 We thus arrive at the following backward relation
 \begin{equation}\label{backward1}
 xM_{\vec{n}}^{\beta +1}(x-1)=M_{\vec{n}+\vec{e}_i}^{\beta }(x)+\frac{c_i}{1-c_i}(\beta +|\vec{n}|)M_{\vec{n}}^{\beta }(x),\quad i=1,\cdots ,r.
 \end{equation}
 Similarly, it is seen that \eqref{ladder1a} amounts to 
 \begin{equation}
 N_{x,\beta }^r \left[ M_{\vec{n}}^{\beta }(x)+\sum_{k=1}^r \frac{n_k}{1-c_k}M_{\vec{n}-\vec{e}_k}^{\beta }(x)\right] = \eta _{x,\beta }N_{x+1,\beta-1}^r M_{\vec{n}}^{\beta -1}(x+1).
 \end{equation}
 The monicity of the polynomials here implies that 
 \begin{equation}
 \eta _{x,\beta }=\frac{N_{x,\beta }^r}{N_{x+1,\beta-1}^r},
 \end{equation}
 thereby giving the following forward relation:
 \begin{equation}\label{forward1}
 M_{\vec{n}}^{\beta -1}(x+1)=M_{\vec{n}}^{\beta }(x)+\sum_{k=1}^r\frac{n_k}{1-c_k}M_{\vec{n}-\vec{e}_k}^{\beta }(x).
 \end{equation}

\section{Intertwiners and contiguity relations}
Contiguity relations can be obtained in a fashion similar to the way the step relations were derived in the last section. One observes that the operators 
\begin{equation}
\hat{X}_i=a_i+\frac{1}{1-c_i}(\beta +H_0),\quad i=1,\cdots ,r
\end{equation}
and 
\begin{equation}
\hat{Y}_i=1+\sum_{k=1}^r\frac{c_k}{1-c_k}a_k^+
\end{equation}
satisfy the following intertwining relations:
\begin{equation}\label{raising2}
H_i^{\beta +1}\hat{X}_j=\hat{X}_jH_i^{\beta }
\end{equation}
and 
\begin{equation}\label{ladder2}
H_i^{\beta -1}\hat{Y}=\hat{Y}H_i^{\beta },
\end{equation}
for $i,j=1,\cdots ,r$.
From \eqref{raising2} and \eqref{ladder2}, we conclude that 
\begin{align}
 \hat{X}_i \left| \left. x,\beta \right> \right> &= \hat{\xi} _{i,x,\beta } \left| \left. x,\beta +1 \right> \right>,\label{raising2a}\\
 \hat{Y} \left| \left. x,\beta \right> \right> &= \hat{\eta} _{x,\beta } \left| \left. x,\beta -1 \right> \right>\label{ladder2a},
\end{align}
with $\hat{\xi }_{i,x,\beta }$ and $\hat{\eta }_{x,\beta }$ proportionality factors again. Note that in contradistinction to \eqref{raising1a} and \eqref{ladder1a}, the eigenvalue $x$ is unscathed in \eqref{raising2a} and \eqref{ladder2a} with $\hat{X}_i$ (resp. $\hat{Y}$) mapping the eigenstate of $H_i^{\beta }$ with eigenvalue $x$ to the eigenstate of $H_i^{\beta +1}$ (resp. $H_i^{\beta -1}$) with the same eigenvalue $x$.

We now proceed as in the last section. Eq. \eqref{raising2a} is seen to imply that
\begin{equation}
 N_{x,\beta}^r\left[ M_{\vec{n}+\vec{e}_i}^{\beta }(x)+\frac{1}{1-c_i}(\beta +|\vec{n}|)M_{\vec{n}}^{\beta }(x)\right] = \hat{\xi}_{i,x,\beta }N_{x-1,\beta +1}^{r} M_{\vec{n}}^{\beta +1}(x).
 \end{equation}
 with monicity forcing 
 \begin{equation}
 \hat{\xi }_{i,x,\beta }=\frac{N_{x,\beta }^r}{N_{x,\beta+1}^r}(x+\hat{a}),
 \end{equation}
 where $\hat{a}$ is a constant. Here also $\hat{a}$ is determined by taking $\vec{n}=\vec{0}$ and one finds $\hat{a}=\beta $. One thus arrives at the following relation
 \begin{equation}\label{backward2}
 (x+\beta )M_{\vec{n}}^{\beta +1}(x)=M_{\vec{n}+\vec{e}_i}^{\beta }(x)+\frac{1}{1-c_i}(\beta +|\vec{n}|)M_{\vec{n}}^{\beta }(x),\quad i=1,\cdots ,r.
 \end{equation}
 A second relation is obtained from \eqref{ladder2a} which entails 
 \begin{equation}\label{step2}
 M_{\vec{n}}^{\beta-1}(x)=M_{\vec{n}}^{\beta }(x)+\sum_{k=1}^r \frac{c_k}{1-c_k}n_k M_{\vec{n}-\vec{e}_k}^{\beta }(x)
 \end{equation}
 with $\hat{\eta }_{x,\beta }$ required to be
 \begin{equation}
 \hat{\eta }_{x,\beta }=\frac{N_{x,\beta }^r}{N_{x,\beta -1}^r}.
 \end{equation}
 
 Here it should be mentioned that combining \eqref{backward1} and \eqref{backward2}, one finds
\begin{equation}\label{raising}
xM_{\vec{n}}^{\beta+1 }(x-1)-c_i(x+\beta )M_{\vec{n}}^{\beta +1}(x) = (1-c_i)M_{\vec{n}+\vec{e}_i}^{\beta }(x),
\end{equation}
for $i=1,\cdots ,r$ and using \eqref{forward1} and \eqref{step2} one gets
\begin{equation}\label{lowering}
M_{\vec{n}}^{\beta }(x+1)-M_{\vec{n}}^{\beta }(x) = \sum_{k=1}^r n_kM_{\vec{n}-\vec{e}_k}^{\beta +1}(x).
\end{equation}
The raising and lowering relations \eqref{raising} and \eqref{lowering} have been shown \cite{Lee} to yield a difference equation for the multiple Meixner polynomials that reads
\begin{align}
\begin{split}
&\prod_{j=1}^r\left[ c_j(x+\beta )-xT_x^{-1}\right]\circ (T_x-I)M_{\vec{n}}^{\beta ,\vec{c}}(x)\\
&=\sum_{k=1}^r (c_k-1)n_k \prod_{j\ne k} \left[ c_j(x+\beta )-xT_x^{-1} \right] M_{\vec{n}}^{\beta ,\vec{c}}(x),
\end{split}
\end{align}
where $I$ is the identity operator and $T_x$ is the shift operator for $x$, i.e. $T_x[f(x)]=f(x+1)$.

 We can also derive  another ``difference'' equation for the multiple Meixner polynomials not in $x$ but in $\beta $ by combining the relations \eqref{backward2} and \eqref{step2}:
 \begin{align}
 \begin{split}
 &\prod_{j=1}^r\left[ (c_j-1)(x+\beta +1) T_{\beta} +(\beta+ |\vec{n}|)\right] \circ (T_{\beta }-I) M_{\vec{n}}^{\beta ,\vec{c}}(x)\\
 &=\sum_{k=1}^r c_kn_k \prod_{j\ne k} \left[ (c_j-1)(x+\beta +1 ) T_{\beta} +(\beta +|\vec{n}|)\right]  M_{\vec{n}}^{\beta +1,\vec{c}}(x),
 \end{split}
 \end{align}
 with $T_{\beta }$ the shift operator for $\beta $.

 \section{Generating function}
 A generating function for the multiple Meixner polynomials of the first kind is readily obtained from the model with the help of the Bargmann realization. With $a_i$ and $a_i^+$ realized as in \eqref{Bargmann}, the Hamiltonian $H_i^{\beta ,\vec{c}}$ reads 
 \begin{equation}
 H_i^{\beta ,\vec{c}}=\sum_{k=1}^r f_{ik}(\vec{z}) \frac{\partial }{\partial z_k}+g_i(\vec{z}),
 \end{equation}
 where
 \begin{align}
 f_{ik}(\vec{z})&= \delta_{ik}+\left( \frac{c_i}{1-c_i}+\frac{1}{1-c_k}\right) z_k +\sum_{j=1}^r \frac{c_j}{(1-c_j)^2} z_jz_k,\\
 g_i(\vec{z})&= \beta \left( \sum_{j=1}^r \frac{c_j}{(1-c_j)^2}z_i+\frac{c_i}{1-c_i}\right) .
 \end{align}
 The eigenvalue equation $H_i^{\beta ,\vec{c}}\left| \left. x,\beta \right> \right> = x\left| \left. x,\beta \right> \right>$ thus becomes a first order linear partial differential equation in $\mathbb{C}^r$. One can check directly that it has for solution the function
 \begin{equation}
 \varphi_x^{\beta ,\vec{c}}(\vec{z})= \left( 1+\sum_{k=1}^r \frac{z_k}{1-c_k}\right) ^x \left( 1+\sum_{k=1}^r \frac{c_k}{1-c_k}z_k\right) ^{-x-\beta }
 \end{equation}
 with the constant of integration chosen so that $\varphi_x^{\beta ,\vec{c}}(0)=1$. Recalling how the number states are represented, see \eqref{Bargmann}, in the Bargmann realization we have:
\begin{equation}
\left< \left. z_1,\cdots ,z_r| x,\beta ,\vec{c}~\right> \right> = N_{x,\beta}^r \sum_{\vec{n}=\vec{0}}^{\infty } \frac{M_{\vec{n}}^{\beta ,\vec{c}}(x)}{n_1!\cdots n_r!}z_1^{n_1}\cdots z_r^{n_r}.
\end{equation} 
Note that $\left< \left. 0,\cdots ,0| x,\beta ,\vec{c}~\right> \right>= N_{x,\beta }^r$; hence, $\varphi_x^{\beta ,\vec{c}}(\vec{z})=\frac{1}{N_{x,\beta}^r} \left< \left. z_1,\cdots ,z_r| x,\beta ,\vec{c}~\right> \right>$.
Bringing this together gives the generating function formula 
\begin{equation}
\left( 1+\sum_{k=1}^r \frac{z_k}{1-c_k}\right) ^x \left( 1+\sum_{k=1}^r \frac{c_k}{1-c_k}z_k\right) ^{-x-\beta }=\sum_{\vec{n}=\vec{0}}^{\infty } \frac{M_{\vec{n}}^{\beta ,\vec{c}}(x)}{n_1!\cdots n_r!}z_1^{n_1}\cdots z_r^{n_r}.
\end{equation}

\section{Conclusions}
Let us summarize our results. As was done previously for the multiple Charlier polynomials, we have provided here an algebraic interpretation of the multiple Meixner polynomials of the first kind. They were shown to arise in the construction of the common eigenfunctions of a set of $r$ non-Hermitian Hamiltonians $H_i^{\beta ,\vec{c}}~~i=1,\cdots ,r$. These turn out to be in involution only ``on shell'' where they define a weakly integrable system. This model is not a multi-variable extension of the familiar connection between the one-dimensional harmonic oscillator and the Meixner polynomials via the metaplectic representation of $SU(1,1)$. The isospectrality of the $H_i$ has been traced back to the fact that their spectra are identical to those of operators $\bar{H}_i$ that can be obtained from the standard $r$-dimensional harmonic oscillator Hamiltonian through conjugation by some specific operators. Finally, the model has been shown to give interpretations to  the step and contiguity relations as well as to a generating function.

\subsection*{Acknowledgement}
One of us (H.M.) wishes to thank the CRM for its hospitality
while this work was carried out and gratefully acknowledges a Grant-in-Aid for Japan Society for the Promotion of Science (JSPS)
Fellows. The research of S.T. is funded in part by KAKENHI (22540224). L.V. is supported in part through funds
provided by the National Sciences and Engineering Research Council (NSERC) of Canada.


\LastPageEnding
\end{document}